\documentclass[twocolumn,preprintnumbers,showpacs,prl,floatfix,amsmath,amssymb,notitlepage,superscriptaddress]{revtex4-1}
\usepackage{epsfig}
\usepackage{subfigure}
\usepackage{amsmath}
\usepackage{color}
\usepackage{amssymb}
\usepackage{setspace}
\usepackage{graphicx}
\usepackage{dcolumn}
\usepackage{bm}
\usepackage{times}
\usepackage{enumerate}
\usepackage{footnote}
\input epsf

\newcommand{\appropto}{\mathrel{\vcenter{
  \offinterlineskip\halign{\hfil$##$\cr
    \propto\cr\noalign{\kern2pt}\sim\cr\noalign{\kern-2pt}}}}}

\graphicspath{{figures/}}

\begin{document}

\author{Per Sebastian Skardal}
\email{skardals@gmail.com} 
\affiliation{Departament d'Enginyeria Informatica i Matem\'{a}tiques, Universitat Rovira i Virgili, 43007 Tarragona, Spain}

\author{Dane Taylor}
\affiliation{Statistical and Applied Mathematical Sciences Institute, Research Triangle Park, NC 27709, USA}
\affiliation{Department of Mathematics, University of North Carolina, Chapel Hill, NC 27599, USA}

\author{Jie Sun}
\affiliation{Department of Mathematics, Clarkson University, Potsdam, NY 13699, USA}

\author{Alex Arenas}
\affiliation{Departament d'Enginyeria Informatica i Matem\'{a}tiques, Universitat Rovira i Virgili, 43007 Tarragona, Spain}

\title{Erosion of synchronization in networks of coupled oscillators}

\begin{abstract}
We report erosion of synchronization in networks of coupled phase oscillators, a phenomenon where perfect phase synchronization is unattainable in steady-state, even in the limit of infinite coupling. An analysis reveals that the total erosion is separable into the product of terms characterizing coupling frustration and structural heterogeneity, both of which amplify erosion. The latter, however, can differ significantly from degree heterogeneity. Finally, we show that erosion is marked by the reorganization of oscillators according to their node degrees rather than their natural frequencies.
\end{abstract}


\maketitle

Synchronization of network-coupled oscillators is a central topic of research in the field of complex systems~\cite{Strogatz2003,Motter2005PRE,Arenas2008PR} due to its importance in many natural \cite{Buck2988QRB,Glass1988} and engineered systems~\cite{Fischer2006PRL,Motter2013NaturePhysics}. In the case of diffusively-coupled limit-cycle oscillators, Kuramoto showed~\cite{Kuramoto1984} that the dynamics of an ensemble of $N$ oscillators can be treated through a reduction to $N$ phase oscillators $\theta_i$ for $i=1,\dots,N$. When placed on a network, the evolution of each oscillator is governed by
\begin{align}
\dot{\theta}_i=\omega_i+K\sum_{j=1}^NA_{ij}H\left(\theta_j-\theta_i\right),\label{eq:Kuramoto}
\end{align}
where $\omega_i$ is the natural frequency of oscillator $i$, $K>0$ is the global coupling strength, $\left[A_{ij}\right]$ is the adjacency matrix encoding the network, and $H$ is the coupling function, which we assume to be $2\pi$-periodic and continuously differentiable. 

Investigations of Eq.~\eqref{eq:Kuramoto} have deepened our understanding of emergent collective behavior and the interplay between structure and dynamics~\cite{Daido1996PRL,Moreno2004EPL,Restrepo2005PRE,Arenas2006PRL,GomezGardenes2007PRL,Gardenes2011PRL,Skardal2011PRE,Skardal2012PRE,Witthaut2014PRE,Skardal2014PRL,Jorg2014PRL}. A key and essential element in this vein is the coupling function $H(\theta)$ that encodes the manner in which interactions between oscillators occur. In particular, we refer to interactions as {\it frustrated} if $h=|H(0)/\sqrt{2}H'(0)|>0$. In other words, a system is frustrated if the contribution to the dynamics in Eq.~(\ref{eq:Kuramoto}) does not vanish when all phases are equal. The presence of coupling frustration is essential in modeling excitable and reaction-diffusion dynamics because neighboring oscillators cannot react simultaneously, but rather one after another~\cite{Shima2004PRE}. Such examples are numerous in biological, chemical, and physical systems including neuron excitation~\cite{FitzHugh1955}, cardiac dynamics~\cite{Karma2007}, and the Belousov-Zhabotinsky reaction~\cite{Winfree1984}. With regard to the network dynamics we ask, what are the consequences of coupling frustration? In this Letter we introduce and study {\it erosion of synchronization}, a novel phenomenon that occurs in frustrated systems in which perfect phase synchronization (characterized by $\theta_1=\theta_2=\dots=\theta_N$) becomes unattainable even in the limit of infinite coupling strength. We  measure the degree of phase synchronization using the order parameter~\cite{Kuramoto1984}
\begin{align}
re^{i\psi} = \frac{1}{N}\sum_{j=1}^Ne^{i\theta_j},\label{eq:r}
\end{align}
where $re^{i\psi}$ denotes the phases' centroid in the complex unit circle and $r$ ranges from $0$ (complete incoherence) to $1$ (perfect synchronization). We define {\it total erosion} as the limiting value of $1-r$ as $K\to\infty$, denoted as $1-r^\infty$. Our analysis reveals that total erosion is separable into a product of two terms characterizing, respectively, the coupling frustration and structural heterogeneity of the network, which relates to the alignment of the nodal degrees with the eigenvectors of the network Laplacian matrix. We find that both coupling frustration and structural heterogeneity amplify erosion. 
Despite the nontrivial dependence of total erosion on network structure, the synchronized oscillators in fact organize according to their degrees instead of their natural frequencies. For the remainder of this Letter we present a general analysis that illustrates our findings, which we then support with numerical experiments.

We now consider the dynamics of Eq.~(\ref{eq:Kuramoto}) in the strong coupling regime where $r\approx1$, following Ref.~\cite{Skardal2014PRL}. We note that such a state can be obtained in several ways, most notably by considering a large enough coupling strength or a small enough spread in the natural frequencies. In fact, these two cases are the same since increasing $K$ is equivalent to appropriately rescaling the natural frequencies and time. Thus, the results presented in this Letter describe the dynamics in both cases. For simplicity we assume that the underlying network is connected, unweighted, and undirected, such that $A_{ij}=1$ if an edge exists between nodes $i$ and $j$, and otherwise $A_{ij}=0$. In the strong coupling regime phases become tightly clustered around the mean phase $\psi$ such that $|\theta_i-\theta_j|\ll1$ for all $(i,j)$ pairs. We will later  numerically corroborate this assumption. In this regime Eq.~(\ref{eq:Kuramoto}) can be linearized to
\begin{align}
\dot{\theta}_i\approx\omega_i+KH(0)d_i-KH'(0)\sum_{j=1}^NL_{ij}\theta_j,\label{eq:Theory01}
\end{align}
where $d_i=\sum_{j=1}^NA_{ij}$ is the degree of node $i$ and $[L_{ij}]$ is the Laplacian matrix, defined by $L_{ij}=\delta_{ij}d_i-A_{ij}$, which has the following spectral properties. First, since the network is connected and undirected, all eigenvalues of $L$ are real, non-negative, and can be ordered $0=\lambda_1 <\lambda_2\le\dots\le\lambda_{N-1}\le\lambda_N$. Second, the normalized eigenvectors $\{\bm{v}^i\}_{i=1}^N$ form an orthonormal basis for $\mathbb{R}^N$. The eigenvector associated with $\lambda_1=0$ is $\bm{v}^1\propto\bm{1}$, which corresponds to the synchronization manifold. Finally, $L$ has pseudo-inverse $L^\dagger=\sum_{j=2}^N\lambda_j^{-1}\bm{v}^j\bm{v}^{j T}$~\cite{BenIsrael1974}, whose null space is the span of $\bm{v}^1$. Thus, $L^\dagger$ projects vectors onto the subspace of zero-mean vectors. 

Inspecting Eq.~(\ref{eq:Theory01}), we find that the mean velocity is $\Omega=\langle\omega\rangle + KH(0)\langle d\rangle$, where $\langle\cdot\rangle$ denotes average over the population. In the rotating reference frame $\theta_i\mapsto\theta_i-\Omega t$, the steady-state solution is given by
\begin{align}
\bm{\theta}^* =\frac{L^\dagger}{H'(0)}\left(\frac{\bm{\omega}}{K}+H(0)\bm{d}\right),\label{eq:Theory02}
\end{align}
where $\bm{\omega}$ and $\bm{d}$ respectively denote the vectors containing the node frequencies and degrees. We now consider the order parameter for the steady-state solution $\bm{\theta}^*$ in Eq.~(\ref{eq:Theory02}). First, in the strong coupling regime, $|\theta_j^*|\ll1$ for all $j$, and thus Eq.~(\ref{eq:r}) can be rewritten as
\begin{align}
r\simeq1-\|\bm{\theta}^*\|^2/2N.\label{eq:Theory03}
\end{align}
Next, by the spectral decomposition of the pseudo-inverse $L^\dagger$ and using $\|\bm{\theta}^*\|^2=\langle\bm{\theta}^*,\bm{\theta}^*\rangle$, we obtain
\begin{align}
r\simeq1-J(\bm{\tilde{\omega}},L)/2K^2H'^2(0),\label{eq:Theory05}
\end{align}
where $\bm{\tilde{\omega}}=\bm{\omega}+KH(0)\bm{d}$ and $J$ is the {\it synchrony alignment function}~\cite{Skardal2014PRL}
\begin{align}
J(\bm{\tilde{\omega}},L)=\frac{1}{N}\sum_{j=2}^N\lambda_j^{-2}\langle\bm{v}^j,\bm{\tilde{\omega}}\rangle^2.\label{eq:Theory06}
\end{align}
In Ref.~\cite{Skardal2014PRL} we derived the synchrony alignment function to optimize synchronization properties of networks under the condition that $H(0)\ll H'(0)$. Here, we will demonstrate its utility in studing erosion of synchronization. In particular, in the limit $K\to\infty$ we obtain from Eqs.~\eqref{eq:Theory05} and~\eqref{eq:Theory06} that
\begin{align}
1-r^\infty\simeq\frac{H^2(0)}{2H'^2(0)}J(\bm{d},L).\label{eq:Theory07}
\end{align}

Equation~\eqref{eq:Theory07} provides a quantitative measure of the total erosion of synchronization as a product of the square of the coupling frustration $h=|H(0)/\sqrt{2}H(0)|$ and the structural heterogeneity $J(\bm{d},L)$. Note that the natural frequencies $\bm{\omega}$ have no effect on the total erosion. We point out that this separation allows us to seamlessly combine the coupling and structural properties of the network to predict the total erosion of synchronization and is reminiscent of the separation of dynamical and structural properties in the Master Stability Function approach for analyzing network synchronization of identical \cite{Pecora1998PRL} and nearly-identical oscillators~\cite{Sun2009EPL}. Equation~\eqref{eq:Theory07} also implies that perfect synchronization is possible if and only if $h$ or $J(\bm{d},L)$ is zero. For a fixed network with non-zero $J(\bm{d},L)$, the total erosion is amplified by coupling frustration and disappears only in its absence. On the other hand, for a given coupling function, the effect on network structure on erosion may be understood through $J(\bm{d},L)$. While it is straightforward to show that $J(\bm{d},L)=0$ for regular networks (i.e., $d_1=d_2=\dots=d_N$), as we will show with numerical experiments, increasing the amount of degree heterogeneity can either increase or decrease $J(\bm{d},L)$.

Two key observations follow from the theory developed in Eqs.~\eqref{eq:Theory01}-\eqref{eq:Theory07}. First, a tightly clustered state $|\theta_i-\theta_j|\ll1$ for all $(i,j)$ pairs is equivalent to $r$ being close to one. From Eq.~(\ref{eq:Theory07}) it follows that our theory remains valid provided that $h^2J(\bm{d},L)\ll1$, which we demonstrate in examples below. Second, we note that the steady-state solution in Eq.~(\ref{eq:Theory02}) is stable for small enough $h$ since the Jacobian $DF$ of Eq.~(\ref{eq:Kuramoto}) evaluated at $\bm{\theta}^*$ is approximately proportional to the negative Laplacian, and thus its spectrum is contained in the left-half complex plane. Towards the end of the Letter we will demonstrate that the solution remains stable for significantly larger frustration values as well.

\begin{figure}[t]
\centering
\epsfig{file =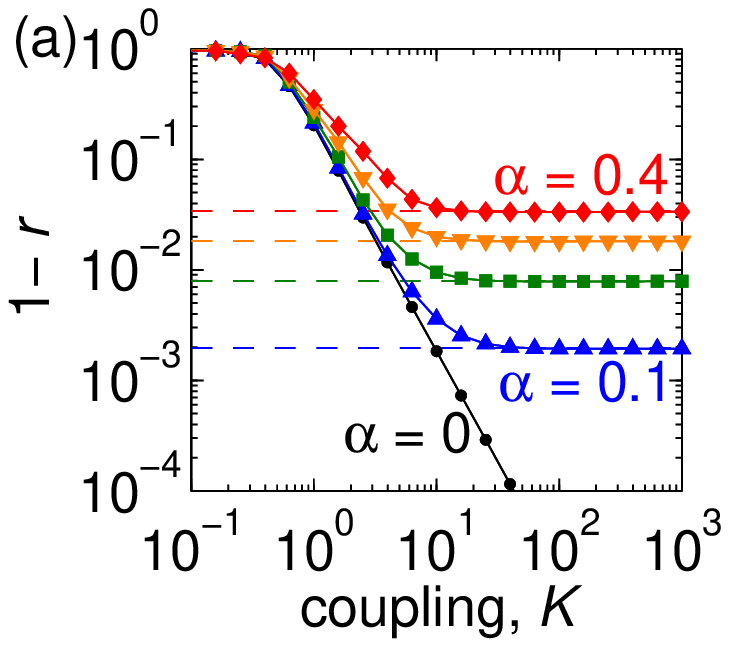, clip =,width=0.49\linewidth } 
\epsfig{file =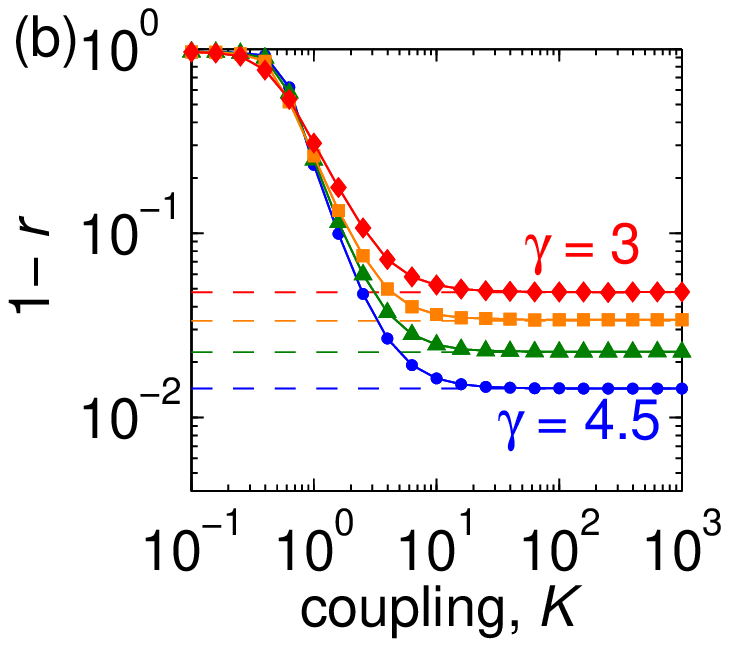, clip =,width=0.49\linewidth } 
\epsfig{file =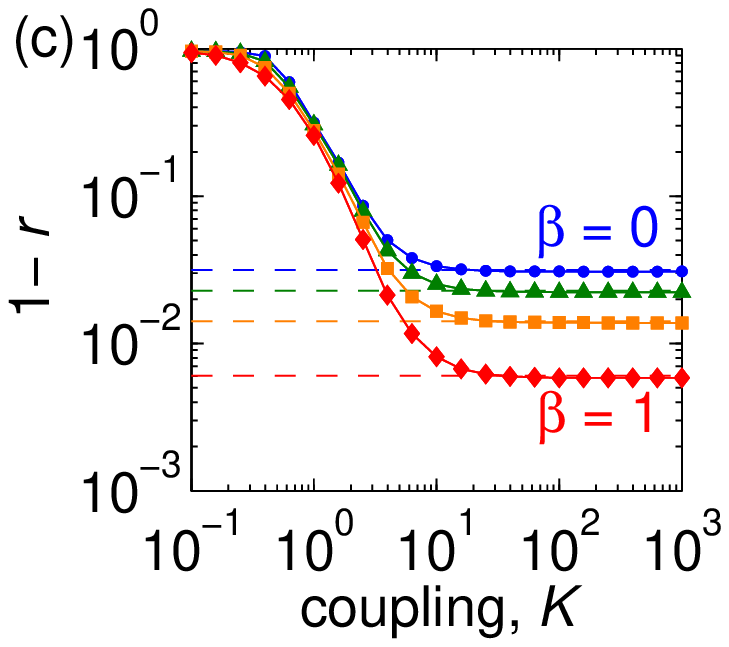, clip =,width=0.49\linewidth } 
\epsfig{file =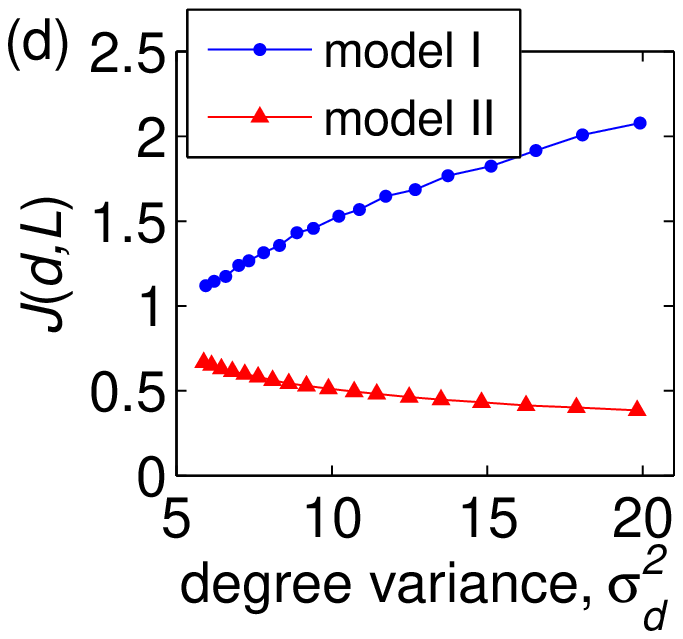, clip =,width=0.49\linewidth } 
\caption{(Color online) \textit{Erosion of synchronization.} Profiles $1-r$ vs $K$ for (a) a fixed network of model I ($\gamma = 3$) varying $\alpha$ between $0$ (black circles, bottom) and $0.4$ (red diamonds, top), (b) several networks of model I with $\gamma$ between $4.5$ (blue circles, bottom), and $3$ (red diamonds, top) with fixed $\alpha=0.2$, and (c) several networks of model II with $\beta$ between $0$ (blue circles, top) and $1$ (red diamonds, bottom) with fixed $\alpha=0.3$. Theoretical predictions for $1-r^\infty$ are denoted with dashed horizontal lines. Other network parameters are $N=1000$, $\langle d\rangle=4$, and $m=1.6$. (d) $J(\bm{d},L)$ vs degree variance $\sigma_d^2$ for network models I (blue circles) and II (red triangles). Data point in (b) and (c) represent an averages over $20$ networks, and data point in (d) represent averages over $100$ networks.}
\label{fig1}
\end{figure}

As an illustrative example of our theory, we consider for the remainder of this Letter the Sakaguchi-Kuramoto (SK) model~\cite{Sakaguchi1986PTP}, which is characterized by $H(\theta)=\sin(\theta-\alpha)$ and has found extensive applications in reaction-diffusion~\cite{Kuramoto1984} and excitable systems~\cite{Kopell2002} and has been linked with the formation of chimera states~\cite{Abrams2004PRL,Abrams2008PRL} and non-universal synchronization transitions~\cite{Omelchenko2012PRL}. Importantly, the coupling frustration $h=|\tan(-\alpha)|/\sqrt{2}$ is tunable via the phase-lag parameter $\alpha\in(-\pi/2,\pi/2)$. 

Moreover, we consider two network models. Model~I consists of scale-free (SF) networks with power-law degree distribution $P(d)\propto d^{-\gamma}$ built using the configuration model with a fixed mean degree $\langle d\rangle$ and tunable exponent $\gamma$~\cite{Molloy1995}. Model~II is given by the following generalization of Ref.~\cite{GomezGardenes2006PRE}: For a prescribed heterogeneity parameter $\beta\in[0,1]$ and minimum degree $d_0$, a network is initialized with $d_0+1$ fully connected nodes. Nodes are then added one-by-one, each making $d_0$ links to the previously existing nodes until the network consists of a total of $N$ nodes. Each link is made either preferentially or at random: with probability $\beta$ the link is made preferentially to a node $i$ with probability $p_i\propto(d_i-m)$, and otherwise the link is made uniformly at random. Here the parameter $m<d_0$ modifies the heterogeneity of the network. Networks generated by model II have mean degree $\langle d\rangle=2d_0$. When $\beta=0$ the model yields an Erd\H{o}s-R\'{e}nyi random network~\cite{Erdos1960} with an approximately Poisson degree distribution, whereas $\beta=1$ gives a preferential attachment network~\cite{Barabasi1999Science} with degree distribution $P(d)\propto d^{-\gamma}$ with $\gamma=3-m/d_0$. In all simulations the natural frequencies are independently drawn from the unit normal distribution.

We now numerically explore erosion of synchronization. Beginning with the effect of frustration, we consider for model~I a fixed network of size $N=1000$ with $\gamma=3$, $\langle d\rangle=4$, and varying $\alpha$. In Fig.~\ref{fig1}(a) we show $1-r$ as a function of the coupling strength $K$ from simulations of Eq.~(\ref{eq:Kuramoto}) using $\alpha$ values between $0$ to $0.4$. While $1-r$ decays as a power-law for $\alpha=0$ (i.e., no frustration), for non-zero $\alpha$ the $1-r$ values approach their expected values given by Eq.~(\ref{eq:Theory07}) (dashed horizontal lines).

To explore the effect of network structure on erosion we first consider networks from model I with $N=1000$, $\langle d\rangle=4$, varying $\gamma$, and fixed $\alpha=0.3$. As shown in Fig.~\ref{fig1}(b), for each $\gamma$ the value $1-r$ approaches $1-r^\infty$ as the coupling strength $K$ increases. The total erosion $1-r^\infty$ increases as $\gamma$ decreases, that is, increased degree heterogeneity amplifies the total erosion. Interestingly, the same does not hold true if we consider networks of model~II. Figure~\ref{fig1}(c) shows that for model~II networks of the same size $N=1000$ and average degree $\langle d\rangle=4$, total erosion is in fact mitigated by increasing the degree heterogeneity, here represented by the increase of $\beta$. This surprising result suggests that erosion of synchronization depends significantly on other microscopic and macroscopic properties of the network in a highly nontrivial manner. This is further supported in Fig.~\ref{fig1}(d), where we show that $J(\bm{d},L)$ tends to increase and decrease, respectively, in network models~I and II as the degree variance $\sigma_d^2=\langle d^2\rangle-\langle d\rangle^2$ increases.

\begin{figure}[t]
\centering
\epsfig{file =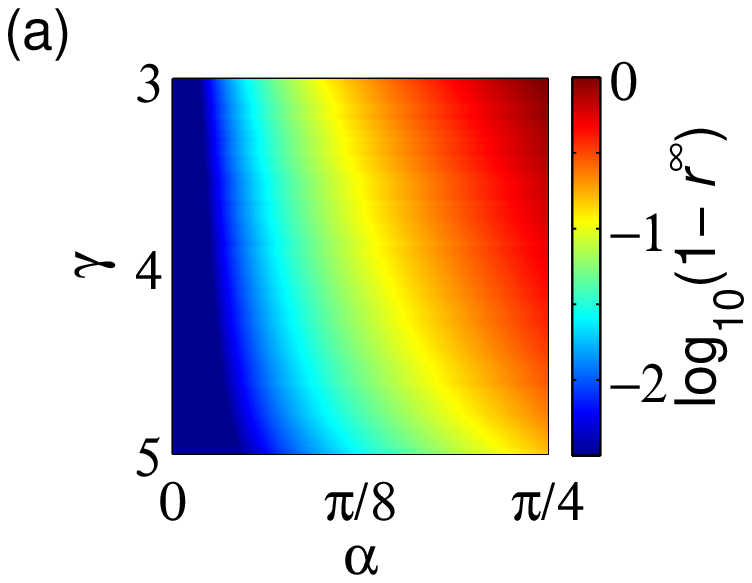, clip =,width=0.49\linewidth } 
\epsfig{file =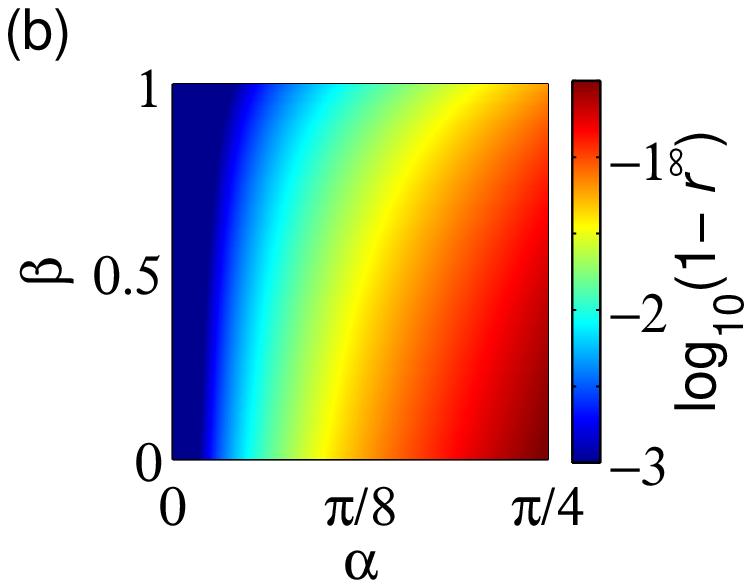, clip =,width=0.49\linewidth } 
\caption{(Color online) \textit{Parameter space of erosion.} Total erosion $1-r^\infty$ as a function of heterogeneity ($\gamma$ or $\beta$) and phase-lag $\alpha$ for (a) model~I and (b) model~II. Results are averaged over $100$ network realizations with $N=1000$, $\langle d\rangle=4$, and $m=1.6$.} \label{fig2}
\end{figure}

To explore the combined effect of coupling frustration and network structure on the erosion of synchronization, we present in Fig.~\ref{fig2}(a) and (b), respectively, the parameter space for the total erosion of synchronization for models~I and II. Specifically, color indicates $1-r^\infty$ as a function of both frustration (parameter $\alpha$) and network heterogeneity (parameter $\gamma$ in model I and $\beta$ in model II), which once again highlights that while greater frustration amplifies erosion, larger degree heterogeneity either amplifies (model~I) or mitigates (model~II) erosion.

\begin{figure}[b]
\centering
\epsfig{file =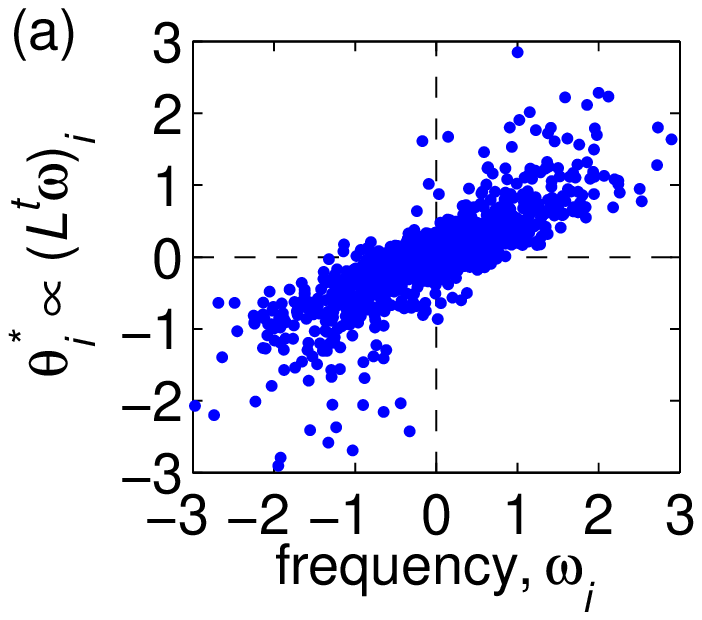, clip =,width=0.49\linewidth } 
\epsfig{file =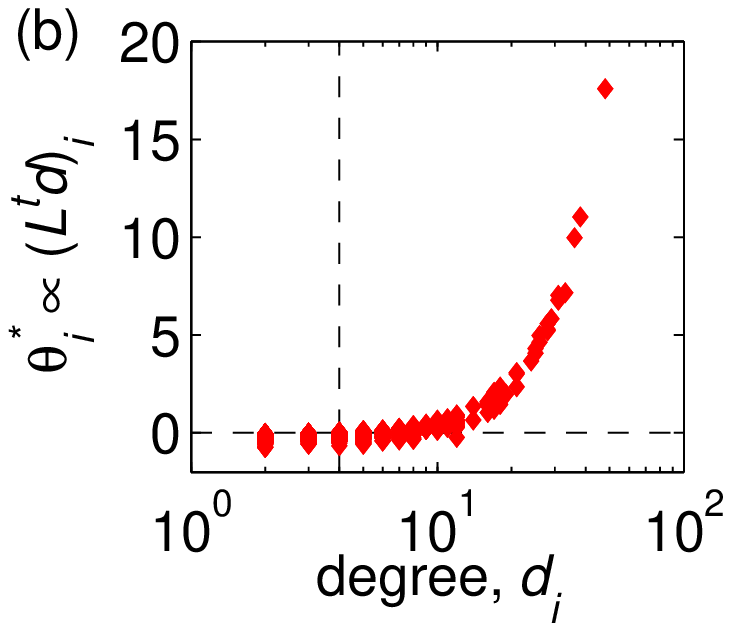, clip =,width=0.49\linewidth } 
\caption{(Color online) \textit{Microscopic reorganization.} Organization in a synchronized state (a) without frustration $\theta_i^*\propto(L^\dagger\bm{\omega})_i$ vs $\omega_i$ and (b) with frustration $\theta_i^*\propto(L^\dagger\bm{\omega})_i$ vs $\omega_i$. The network is model~II with $N=1000$, $\beta=0.5$, $\langle d\rangle=4$, and $m=1.6$. For easy visualization we have normalized $\bm{\theta}^*$ to have unit standard deviation.}\label{fig:Linv}
\end{figure}

Next we discuss the microscopic properties of synchronized states, which will elucidate the mechanism for erosion of synchronization. Revisiting Eq.~(\ref{eq:Theory01}), the two contributing terms to the heterogeneity of the oscillators' dynamics are the natural frequencies $\omega_i$ and the degrees $KH(0)d_i$. When the coupling frustration $h=|H(0)/\sqrt{2}H'(0)|=0$ the latter vanishes and heterogeneity is captured solely in the natural frequencies. Thus, from Eq.~(\ref{eq:Theory02}) we see that in a strongly synchronized state oscillators organize themselves according to $\bm{\theta^*}\propto L^\dagger\bm{\omega}$, i.e., their positions are determined jointly by the network structure as well as natural frequencies. In fact, the oscillators' organization has a strong positive correlation with the natural frequencies, which we illustrate in Fig.~\ref{fig:Linv}(a) for a network of type II ($N=1000$, $\beta=0.5$, $\langle d\rangle=4$, $m=1.6$). On the other hand, in the presence of frustration ($h\neq0$), for large enough $K$ the term $KH(0)d_i$ dominates and the natural frequencies can be neglected. In this case Eq.~(\ref{eq:Theory02}) implies that oscillators in a strongly synchronized state organize according to $\bm{\theta^*}\propto cL^\dagger\bm{d}$, where the sign of $c$ is determined by $H(0)/H'(0)$, i.e., their position depends solely on the network structure. The oscillators' organization is strongly correlated with the degrees, which we illustrate in Fig.~\ref{fig:Linv}(b).

\begin{figure}[t]
\centering
\epsfig{file =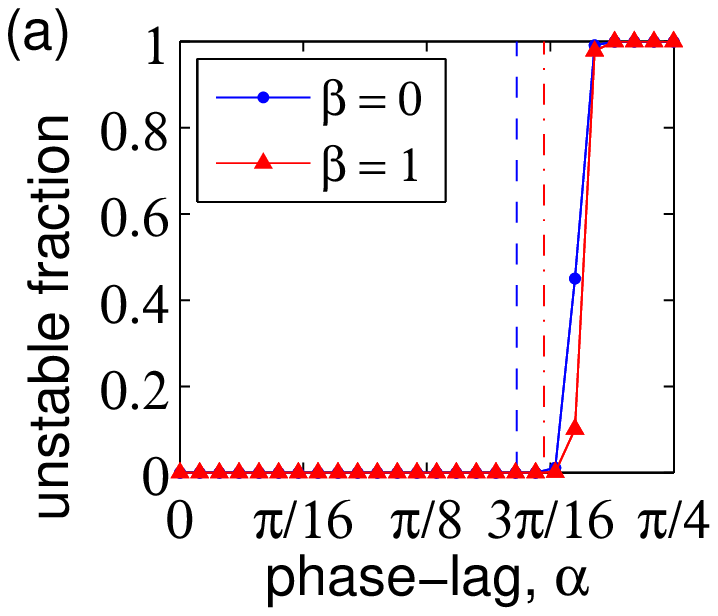, clip =,width=0.49\linewidth } 
\epsfig{file =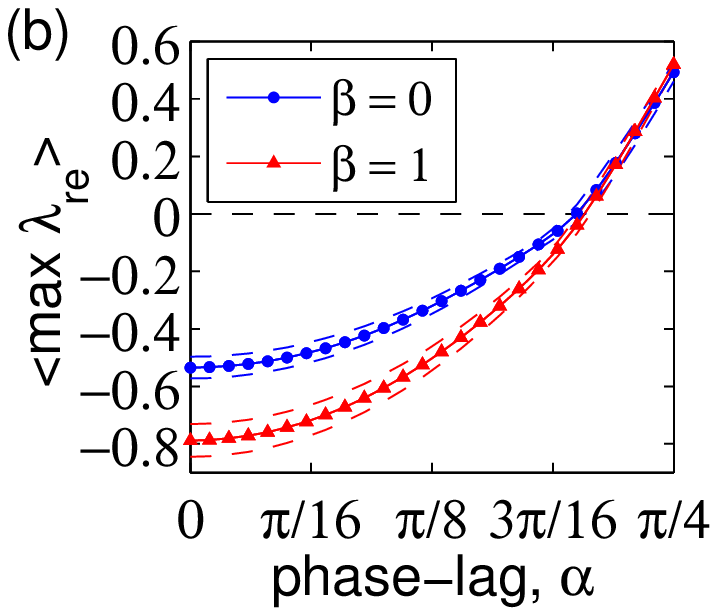, clip =,width=0.49\linewidth } 
\caption{(Color online) \textit{Stability of synchronization.} Averaged over $1000$ model II networks with $\beta=0$ (blue circles) and $1$ (red triangles), (a) the fraction of unstable solutions and (b) the average real part of  the maximal nontrivial Jacobian eigenvalues (standard deviation indicated by dashed line tubes). Other network parameters are $N=1000$, $\langle d\rangle=4$, and $m=1.6$. Each data point is an average over $1000$ independent network realizations.
}\label{fig:Stability}
\end{figure}

Before concluding, we briefly discuss the stability of the steady-state solution $\bm{\theta}^*$ given by Eq.~(\ref{eq:Theory02}). In particular, $\bm{\theta}^*$ is stable if the nontrivial eigenvalues of the Jacobian $DF$ of Eq.~(\ref{eq:Kuramoto}) all have negative real part. The entries of the Jacobian are given by $DF_{ij} = KA_{ij}H'(\theta_j^*-\theta_i^*)$ for $j\ne i$ and $DF_{ii}=-K\sum_{j\ne i}A_{ij}H'(\theta_j^*-\theta_i^*)$. Thus, for small frustration $DF\appropto-L$ and consequently the solution is stable.  As the coupling frustration increases, the eigenvalues of $DF$ can potentially cross into the right-half complex plane, rendering the solution $\bm{\theta}^*$ unstable. We investigate this transition by computing the spectra of networks from model II with two different choices of $\beta$ (0 and 1) and $\alpha\in[0,\pi/4]$. Other network parameters are $N=1000$, $\langle d\rangle=4$, and $m=1.6$. In Fig.~\ref{fig:Stability}(a) we plot the fraction of 1000 network realizations that yield unstable solutions, and in Fig.~\ref{fig:Stability}(b) we plot the maximum real part of the nontrivial eigenvalues plus and minus the standard deviation (dashed curves). The instability transition occurs only if at least one entry in the off-diagonal of $DF$ is positive, which gives a necessary condition for instability,
\begin{equation}
\min_{A_{ij}\neq0}H'\left(\frac{H(0)}{H'(0)}\Big[(L^\dagger\bm{d})_j-(L^\dagger\bm{d})_i\Big]\right)<0.\label{eq:instability}
\end{equation}
Using Eq.~(\ref{eq:instability}) we calculate for each experiment above the lower-bound $\alpha_c$ for instability to occur, indicating the average $\alpha_c$ with vertical lines (dashed and dot-dashed for $\beta=0$ and $1$, respectively) in Fig.~\ref{fig:Stability}(a), which are in good agreement with our numerics.

In this Letter we have investigated erosion of synchronization, a novel phenomenon where perfect synchronization becomes unattainable in steady-state, even in the limit of infinite coupling strength. Our analysis reveals that erosion arises due to the presence of two system properties: frustration in the coupling function governing the oscillators' interactions and structural heterogeneity of the underlying network. In particular, the total erosion of synchronization can be quantified as a product of terms that correspond to these respective properties, and which both amplify erosion. Erosion is marked by oscillators reorganizing themselves according to their local network structure, rather than according to their natural frequencies. Finally, we showed that a sufficiently large amount of frustration can cause the synchronized state to lose stability.

Our theoretical results center on the synchrony alignment function given by Eq.~(\ref{eq:Theory06}), which is a quantitative measure of the interplay between a vector (here the degree vector $\bm{d}$) and the network Laplacian. The synchrony alignment function was recently derived and utilized for the optimization of synchronization in the absence of coupling frustration~\cite{Skardal2014PRL}. Here, we adopted it as an analytical tool for studying the erosion of synchronization that emerges in networks of coupled oscillators under the presence of coupling frustration. One particularly interesting and somewhat counterintuitive finding is that the structural heterogeneity of a network cannot be merely extrapolated from its degree distribution or other simple local characteristics such as degree-degree correlation and clustering. Depending on the detailed connection of the nodes, a network with a relatively homogeneous degree distribution can in fact be relatively structural heterogeneous, and vice versa. Given the importance of frustrated interactions in physical, chemical, and biological applications \cite{Shima2004PRE,Kopell2002}, deeper investigation into the effects of macroscopic and microscopic network properties could be vital in developing a full understanding of the dynamics of Eq.~(\ref{eq:Kuramoto}) and other coupled systems of nonlinear dynamical systems.

Finally, we point to a possible application and future work. With the goal of optimizing synchronization in mind, one key question is how synchronization erosion can be mitigated when coupling frustration and network heterogeneity are unavoidable. We hypothesize that in a more general setting, coupling frustration and network heterogeneity, both of which individually cause and amplify erosion, can in fact be jointly exploited to improve and even optimize synchronization.

\acknowledgements
This work was funded in part by the James S. McDonnell Foundation (PSS and AA), NSF Grant No. DMS-1127914 through the Statistical and Applied Mathematical Sciences Institute (DT), Simons Foundation Grant No. 318812 (JS), Spanish DGICYT Grant No. FIS2012-38266 (AA), and FwET Project No. MULTIPLEX (317532) (AA).

\bibliographystyle{plain}

\end{document}